\documentclass[aps,prl,twocolumn,superscriptaddress]{revtex4}
\usepackage{graphicx,amssymb}

\begin{document}
\newcommand{\BSCCO}{Bi$_2$Sr$_2$CaCu$_2$O$_{8+\delta}$}

\title{Destruction of antinodal state coherence via `checkerboard' charge ordering in strongly underdoped superconducting \BSCCO}

\author{K. McElroy}
\affiliation{Physics Department, University of California,
Berkeley, CA 94720 USA} \affiliation{Material Sciences Division,
Lawrence Berkeley National Lab., Berkeley, CA 94720  USA}
\affiliation{LASSP, Department of Physics, Cornell University,
Ithaca, NY 14850 USA}
\author{D.-H. Lee}
\affiliation{Physics Department, University of California,
Berkeley, CA 94720 USA}
\affiliation{Material Sciences Division,
Lawrence Berkeley National Lab., Berkeley, CA 94720  USA}
\author{J. E. Hoffman}
\affiliation{Department of Applied Physics, Stanford University,
Stanford, CA 94305, USA}
\author{K. M Lang}
\affiliation{Department of Physics, Colorado College, CO 80903,
USA}
\author{J. Lee}
\affiliation{LASSP, Department of Physics, Cornell University,
Ithaca, NY 14850 USA}
\author{E. W. Hudson}
\affiliation{Department of Physics, MIT, Cambridge MA 02139, USA}
\author{H. Eisaki}
\affiliation{AIST, 1-1-1 Central 2, Umezono, Tsukuba, Ibaraki,
305-8568 Japan}
\author{S. Uchida}
\affiliation{Department of Physics, University of Tokyo, Tokyo,
113-8656 Japan}
\author{J.C. Davis}
\email[]{jcdavis@ccmr.cornell.edu} \affiliation{LASSP, Department
of Physics, Cornell University, Ithaca, NY 14850 USA}

\date{\today}

\begin{abstract}
The doping dependence of nanoscale electronic structure in
superconducting \BSCCO \ is studied by Scanning Tunneling
Microscopy (STM). At all dopings, the low energy density-of-states
modulations are analyzed according to a simple model of
quasiparticle interference and found to be consistent with
Fermi-arc superconductivity. The superconducting coherence-peaks,
ubiquitous in near-optimal tunneling spectra, are destroyed with
strong underdoping and a new spectral type appears. Exclusively in
regions exhibiting this new spectrum, we find local `checkerboard'
charge-order with wavevector $\vec{Q}=(\pm 2\pi/4.5a_0,0)$;
$(0,\pm 2\pi/4.5a_0)\pm 15\%$. Surprisingly, this order coexists
harmoniously with the the low energy Bogoliubov quasiparticle
states.
\end{abstract}

\pacs{71.18.+y}

\maketitle

How the electronic structure evolves with doping from a Mott
insulator into a d-wave superconductor is the key issue in
understanding the cuprate phase diagram. Recently it has become
clear that states in different parts of momentum space exhibit
quite different doping dependences. The Fermi-arc\cite{norman98}
(near nodal) states of superconducting cuprates retain their
coherence as doping is reduced, while the antinodal (at the edge
of the 1$^{{\mathrm st}}$ Brillouin zone) states diminish in
coherence, eventually becoming completely incoherent at strong
underdoping. Photoemission (ARPES) reveals this directly because
the nodal states persist almost into the
insulator\cite{yoshida03,ronning03} while the antinodal states
rapidly become
incoherent\cite{loesser97,fedorov99,feng00,zhou04,ding01}. Bulk
probes like thermal conductivity\cite{sutherland03} and c-axis
penetration depth\cite{hosseini03} also show that Fermi-arc states
survive down to the lowest superconducting dopings. Other probes
such as optical transient grating spectroscopy\cite{gedik03},
Raman scattering\cite{gallais04}, and NMR\cite{pines04} show very
different scattering processes of antinodal versus nodal states
throughout the underdoped regime.

Although there is now agreement on the robust nature of the
Fermi-arc states upon underdoping, a variety of different
mechanisms have been proposed to account for the destruction of
the antinodal coherence. These include strong scattering by both
antiferromagnetic spin fluctuations\cite{pines04,hosseini03} and
between the almost parallel segments of the Fermi surface near the
zone face.\cite{zhou04,fu04}

Here we report on doping-dependent STM studies of \BSCCO (Bi-2212)
to illuminate the mechanism of antinodal decoherence. The local
density of states ($LDOS$) is mapped by measuring the STM
tip-sample differential tunneling conductance
$g(\vec{r},V)\equiv\frac{dI}{dV}|_{r,V}$ at each location
$\vec{r}$ and bias voltage $V$. Since $LDOS(\vec{r},E=eV)\propto
g(\vec{r},V)$, an energy-resolved $\vec{r}$-space electronic
structure map is attained. The magnitude of the energy-gap
$\Delta$, defined as half the energy difference between the
coherence peaks, can also be mapped (gapmap).\cite{lang02}

Fourier transform scanning tunneling spectroscopy (FT-STS) was
recently introduced to cuprate
studies.\cite{hoffman02q,mcelroy03,howald03,hoffman03v,vershinin04}
It allows the $\vec{q}$-vectors of spatial modulations in
$g(\vec{r},V)$ to be determined from the locations of peaks in
$g(\vec{q},V)$, the Fourier transform magnitude of $g(\vec{r},V)$.
This technique has the unique capability to relate the nanoscale
$\vec{r}$-space electronic structure to that in
$\vec{k}$-space.\cite{mcelroy03}

For this study we used single Bi-2212 crystals grown by the
floating zone method with the doping controlled by oxygen
depletion. The samples were cleaved in cryogenic ultra-high vacuum
before immediate insertion into the STM. We acquired more than
$10^6$ spectra for this study.

\begin{table}
  \caption{The various sample doping and average
properties reported here.}\label{samples}
\begin{ruledtabular}
\begin{tabular}{c|c|c|c|c|c|c}
   Fig. 1 & T$_c$ & $p$ (\%) & $\bar{\Delta} (meV)$ & $\sigma$ (meV) & P$_1$& P$_6$ \\
  \hline
  (a) & 89K OD & $19\pm1$ & $33\pm1$ & 7   & 30\% & 0\% \\
  (b) & 79K UD & $15\pm1$ & $43\pm1$ & 9   & 5\%  & 1\% \\
  (c) & 75K UD & $13\pm1$ & $48\pm1$ & 10  & 1\%  & 8\% \\
  (d) & 65K UD & $11\pm1$ & $>62$    & unclear & 0\%  & $>55\%$ \\
\end{tabular}
\end{ruledtabular}
\end{table}

In Fig. 1 we show 50 nm square gapmaps measured on samples with
four different dopings. Identical color scales representing 20 meV
$<\Delta(\vec{r})<70$ meV are used. The local hole concentration
is impossible to determine directly, but we estimate the bulk
dopings in Table \ref{samples}. Above optimal doping (Fig 1(a))
the vast majority of tunneling spectra are consistent with those
of a d$_{x^2-y^2}$ superconductor (see Fig 2(a)). However, at the
lowest dopings and for gap values exceeding $\sim65$ meV, there
are many spectra where $\Delta$ is ill defined because no peaks
exist at the gap edge (e.g. Figure 2(a), spectrum 6). We represent
these spectra by black in gapmaps. The spatially averaged value of
$\Delta(\vec{r})$, $\bar{\Delta}$, and its full width at half
maximum, $\sigma$, are also in Table \ref{samples}. As doping is
reduced, $\bar{\Delta}$ grows steadily consistent with other
spectroscopic techniques, such as ARPES,\cite{damascelli03}
break-junction tunneling,\cite{miyakawa98} and thermal
conductivity\cite{sutherland03} which average over many nanoscale
regions.

In Figure 2(a) we show the average spectrum of all regions
exhibiting a given local gap value. They are color-coded so that
each gap-averaged spectrum can be associated with regions of the
same color in all gapmaps (Fig 1(a)-(d)). This set of gap-averaged
spectra is almost identical for all dopings. The changes with
doping seen in $\Delta(\vec{r})$ occur because the probability of
observing a given type of spectrum (1-6) in Fig 2(a) evolves
rapidly with doping (Table \ref{samples}).

Despite the intense changes with doping in the gapmaps, the $LDOS$
at energies below about 0.5$\bar{\Delta}$ remains relatively
homogenous for all dopings studied [Figure 2(a)]. These low energy
$LDOS$ do, however, exhibit numerous weak, incommensurate,
energy-dispersive, $LDOS$-modulations with long correlation
lengths \cite{hoffman02q,howald03,mcelroy03,vershinin04}. To
explore the doping dependence of these low energy $g(\vec{r},V)$
we use the FT-STS technique and the `octet'
model.\cite{mcelroy03,zhu04} Figure 3(b) (using the
$\vec{q}$-vector designations in Figure 3(a)) shows the measured
length of $\vec{q}_1$, $\vec{q}_5$ and $\vec{q}_7$ as a function
of energy for the three data sets. Figure 3(c) shows the locus of
scattering $\vec{k}_s(E)$\cite{mcelroy03} calculated for these
three $g(\vec{r},V)$. These $\vec{k}_s(E)$ differ only slightly
between dopings and are the same for filled and empty states.

The doping dependence of states with
$\vec{k}\approx(\pm\pi/a_0,0),(0,\pm\pi/a_0)$ in the `flat band'
region near the zone-face (green shaded in Fig 2(a)) is extremely
different. By definition, the coherence peaks in $g(\vec{r},V)$
occur at $E=\Delta(\vec{r})$. In all samples, they exhibit intense
bias symmetric intensity modulations in the $LDOS$, with
wavevectors $\vec{G}=(\pm 2\pi/a_0,0),(0,\pm 2\pi/a_0)$. These
coherence-peak $LDOS$-modulations at $\vec{q}=\vec{G}$ seem to be
from Umpklapp scattering between
$\vec{k}\approx(\pm\pi/a_0,0),(0,\pm\pi/a_0)$.\cite{mcelroy03}
Therefore, the coherence-peaks in tunneling are identified
empirically with the zone-face states at
$\vec{k}\approx(\pm\pi/a_0,0),(0,\pm\pi/a_0)$, consistent with
theory. We therefore consider any spatial regions that show clear
coherence-peaks plus $\vec{q}=\vec{G}$ $LDOS$-modulations to be
occupied by a canonical d-wave superconductor (dSC).

Near optimal doping, more than 98\% of any FOV exhibits this type
of coherence-peaked spectrum. As the range of local values of
$\Delta(\vec{r})$ increases with decreasing doping, the intensity
of the $\vec{q}=\vec{G}$ coherence-peak $LDOS$-modulations becomes
steadily weaker until, wherever $\Delta(\vec{r}) \nless 65$ meV,
they disappear altogether. This process can be seen clearly in the
gap-averaged spectra of Fig 2(a). Wherever the coherence-peaks and
associated $\vec{q}=\vec{G}$ $LDOS$-modulations are absent, a
well-defined new type of spectrum is always observed. Examples of
this new type of spectrum, along with those of coherence-peaked
dSC spectra, are shown in Fig 2(b). The coherence-peaked spectra
(red) are manifestly distinct from the novel spectra (black) which
have a V-shaped gap reaching up to $\pm75$ meV but with very
different evolutions at opposite bias beyond these energies. For
reasons to be discussed below, we refer to the new spectrum (black
in Fig 2(b)) as the zero temperature pseudogap (ZTPG) spectrum.

The replacement of coherence-peaked spectra by ZTPG spectra first
begins to have strong impact on averaged properties of
$g(\vec{r},V)$ and $g(\vec{q},V)$ below about $p=0.13$ where the
fractional area covered by ZTPG spectra first exceeds
$\approx10\%$ of the FOV. No further evolution in spectral shape
of the ZTPG spectrum is detected at lower dopings. Instead, a
steadily increasing fractional coverage of the surface by these
ZTPG spectra is observed. Our previous
studies\cite{lang02,hoffman02q,mcelroy03} were carried out at
dopings $p>0.13$ where ZTPG spectra comprise a tiny fraction of
any FOV. Significantly, spectra similar to the ZTPG spectrum are
detected inside vortex cores of Bi-2212 where superconductivity is
destroyed.\cite{renner98,pan00} Furthermore, a very similar
spectrum is observed in another very underdoped cuprate
Na$_x$Ca$_{2-x}$CuO$_2$Cl$_2$, even in the non-superconducting
phase.\cite{hanaguri04} It therefore seems reasonable that this
spectrum is characteristic of the electronic phase that exists at
zero-temperature in the pseudogap (ZTPG).

Next we introduce a masking process illuminate the electronic
structure of ZTPG nano-regions. From a given strongly underdoped
data set (Fig 1(d)), the $g(\vec{r},V)$ in all regions where
$\Delta(\vec{r}) \nless 65$ meV are excised and used to form a new
masked data set $g(\vec{r},V)|_{\Delta \nless 65}$. The remainder
forms a second new data set $g(\vec{r},V)|_{\Delta<65}$. The
cutoff $\Delta(\vec{r}) < 65$ meV was chosen because it represents
where the coherence-peaks with $\vec{q}=\vec{G}$ associated
modulations have disappeared and are replaced by the ZTPG spectra.

FT-STS analysis of such ($g(\vec{r},V)|_{\Delta \nless
65}$,$g(\vec{r},V)|_{\Delta<65}$) pairs shows that they exhibit
dramatically different phenomena. In the
$g(\vec{r},V)|_{\Delta<65}$, the dispersive trajectory of
$\vec{q}_1$ is seen up to $E\approx36$ meV and no further
$LDOS$-modulations can be detected at any higher energy (red
symbols in Figure 4(b)). In the $g(\vec{r},V)|_{\Delta \nless 65}$
data, the identical dispersive $\vec{q}_1$ signal is seen below
E$\approx36$ meV but, in addition, a new nondispersive
$LDOS$-modulation appears between $E=-65$ meV and our maximum
measured energy E$=-150$ meV (black symbols in Fig 4(a)) with a
wavevector
$\vec{q}^{*}=(\pm2\pi/4.5a_0,0),(0,\pm2\pi/4.5a_0)\pm15\%$.

To search for non-dispersive $LDOS$ modulations integration over
energy is often used.\cite{hoffman03v,howald03} In Figure 4(b) we
show the Fourier transform magnitude of the both components of the
masked $LDOS$ integrated over the energy range where we see
$\vec{q}^{*}$ (from Fig 1 (d)). The ZTPG regions show a peak at
the same well defined wavector set
$\vec{q}^{*}=(\pm2\pi/4.5a_0,0),(0,\pm2\pi/4.5a_0)\pm15\%$ while
the dSC regions show no such effect.

An even more conclusive technique for detection of net charge
density modulations is constant-current topography because it
represents, albeit logarithmically, the contour of constant
integrated density of states. We apply the identical mask ($\Delta
\nless 65$ meV) to the topographic image acquired simultaneously
with the gapmap in Fig 1(d). The magnitude of the Fourier
transform along the $\vec{q}||(2\pi,0)$ for this masked
topographic image shows that, in the ZTPG regions, the topography
is modulated with
$q_{topo}=(\pm2\pi/4.7a_0,0),(0,\pm2\pi/4.7a_0)\pm20\%$ (indicated
by the arrow in Fig 4(c)). No such modulations at any wavelengths
near this $q_{topo}$  are found in $\Delta<65$ meV regions (red in
Fig 4(c)).

A static charge modulation with wavevector $\vec{Q}$
should\cite{vojta02,podolsky03,chen03yeh,capriotti03} influence
quasiparticle scattering to cause an enhancement in $g(\vec{Q},V)$
for any $V$. Our measurements of the intensity of the dispersive
quasi-particle branch, $\vec{q}_{1}$, reveal a maximum in
$g(\vec{q}_1 ,V)$ (in the ZTPG regions only) when $\vec{q}_1
=\vec{Q}=(\pm 2 \pi/4.2a_0,0),(0,\pm 2 \pi/4.2a_0)\pm 15\%$ (in
Fig 4(d)). This provides further evidence of charge order
exclusively in ZPTG regions.\cite{related}

These observations motivate three new insights into the electronic
structure of \BSCCO. First, quasiparticle interference occurs
between states in approximately the same region of $\vec{k}$-space
for all dopings. These Fermi-arc quasiparticles are
Bogoliubov-like in the sense that they exhibit particle-hole
symmetry at each location in $\vec{k}$-space and are consistent
with one $\Delta(\vec{k})$ for each doping. Therefore, the
Fermi-arc states are robust and gapped by superconductivity at all
dopings studied.

Our second finding is the very different fate of states in the
flat-band regions near
$\vec{k}\approx(\pm\pi/a_0,0)(0,\pm\pi/a_0)$. The appearance of
ZTPG spectra in strongly underdoped samples coincides with the
destruction of antinodal superconducting coherence-peaks.
Exclusively in these ZTPG regions, which only occur in strongly
underdoped Bi-2212, multiple independent phenomena with the same
wavevector
$\vec{Q}=\vec{q}_{topo}=\vec{q}^{*}=(\pm2\pi/4.5a_0,0),(0,\pm2\pi/4.5a_0)\pm15\%$
 point to the appearance of an unusual
charge ordered state. We refer to this as a `checkerboard' state
because Fourier analysis of all these phenomena shows that they
are symmetric under 90 rotations.

The third point is that the low energy quasiparticle interference
effects indicate the existence of Bogoliubov quasiparticles in
\textit{both} the ZTPG and dSC regions.

Consistent with previous
deductions\cite{norman98,yoshida03,ronning03,loesser97,fedorov99,feng00,ding01,zhou04,sutherland03,hosseini03,gedik03,gallais04,pines04},
we therefore find two distinct regions of $\vec{k}$-space; the
Fermi-arc supporting robust coherent quasiparticle states, and the
antinodal region gapped by superconductivity at higher doping but
becoming progressively incoherent below $p\sim$0.13. But here we
demonstrate for the first time that antinodal decoherence is
closely related to the emergence of the `checkerboard' charge
order. Furthermore, we find that this order is not mutually
exclusive with the Bogoliubov  Fermi-arc states, but rather they
coexist throughout the sample. This miscibility between the low
energy excitations of a high T$_c$ superconductor and the charge
ordered state has strong implications for the phase diagram. It
also provides severe constraints on microscopic theories of
strongly underdoped cuprate electronic structure.

\textbf{Fig. 1} (a)-(d) Measured $\Delta(\vec{r})$, gapmaps, for
four different hole-doping levels as listed in \ref{samples}.

\textbf{Fig. 2} (a) The average spectrum, $g(E)$, associated with
each gap value in a given FOV from Fig 1. These were extracted
from Fig 1(b) but the equivalent analysis for $g(\vec{r},V)$ at
all dopings yields results which are indistinguishable. The
coherence peaks can be detected in spectra 1-4. (b) Characteristic
spectra from the two regions $\Delta<65$ (red) and $\Delta \nless
65$ (black)

\textbf{Fig. 3}(a) A schematic representation of the 1st Brillouin
zone and Fermi arc location of Bi-2212. The flat-band regions near
the zone face are shaded in green. The eight locations which
determine the scattering within the `octet' model\cite{mcelroy03}
(for one sub gap energy) are show as red circles and the
scattering vectors which connect these locations are show as
arrows labelled by the designation of each scattering vector. (b)
Measured dispersions of the LDOS-modulations $\vec{q}_1$,
$\vec{q}_5$ and $\vec{q}_7$ for the 3 dopings whose gapmaps are
shown in Fig 1 (a), (c), and (d). (c). Calculated loci of
scattering, $\vec{k}_s$ for all 3 dopings.

\textbf{Fig. 4}(a) Dispersion of $\vec{q}_1$ in regions with dSC
coherence-peaked spectra $\Delta<65$ meV (red circles). Dispersion
of in regions with ZTPG spectra for E$< 36$ meV (black squares).
For E$>65$ meV, the wavevector of the $\vec{q}^*$ new modulations
in ZTPG regions are shown in black. To within our uncertainty they
do not disperse and exhibit
$\vec{q}^{*}=(\pm2\pi/4.5a_0,0),(0,\pm2\pi/4.5a_0)\pm15\%$. (b)The
magnitude of the $g(\vec{q},V)$ integrated between -65 meV and
+150 meV along the $\vec{q}||(\pi,0)$ direction for $\Delta \nless
65$ meV ($\Delta<65$ meV) black (red). Solid lines are guides to
the eye.(c) The magnitude of the Fourier transform of the masked
topographic image along the $\vec{q}||(\pi,0)$ direction for
$\Delta \nless 65$ meV and ($\Delta<65$ meV) black (red). Solid
lines are guides to the eye.(d) A plot of the amplitude of the
$\vec{q}_1$ $LDOS$-modulation as a function of $|\vec{q}_1|$ for
the same data set yielding Figure 1(a). The maximum intensity of
the modulations in the ZTPG regions occurs at
$|\vec{q}_1|=2\pi/4.5a0\pm10\%$. No enhanced scattering of the
quasiparticles in the dSC regions (red) is seen near any
$\vec{q}$-vector. The $\vec{q}$-space resolution a masked data set
is far less than that in previous studies.\cite{mcelroy03}

\begin{acknowledgments}
We acknowledge and thank A. V. Balatsky, S. Chakravarty, M. P. A.
Fisher, T. Hanaguri, N. E. Hussey, S. Kivelson, P. A. Lee, A.
Millis, D. Pines, S. Sachdev, J. Sethna, T. Uemura, A. Yazdani, J.
Zaanen, and S.-C. Zhang  for very helpful discussions and
communications.
\end{acknowledgments}


\begin{thebibliography}{99}
\bibitem{related}Similar phenonema have been discovered where superconductivity is destroyed
    leaving pseudogap like spectra. Incommensurate, low energy,
    `checkerboard', $LDOS$-modulations are found both surrounding
    vortex cores\cite{hoffman03v} and in the pseudogap phase above
    T$_c$\cite{vershinin04}
\end{thebibliography}
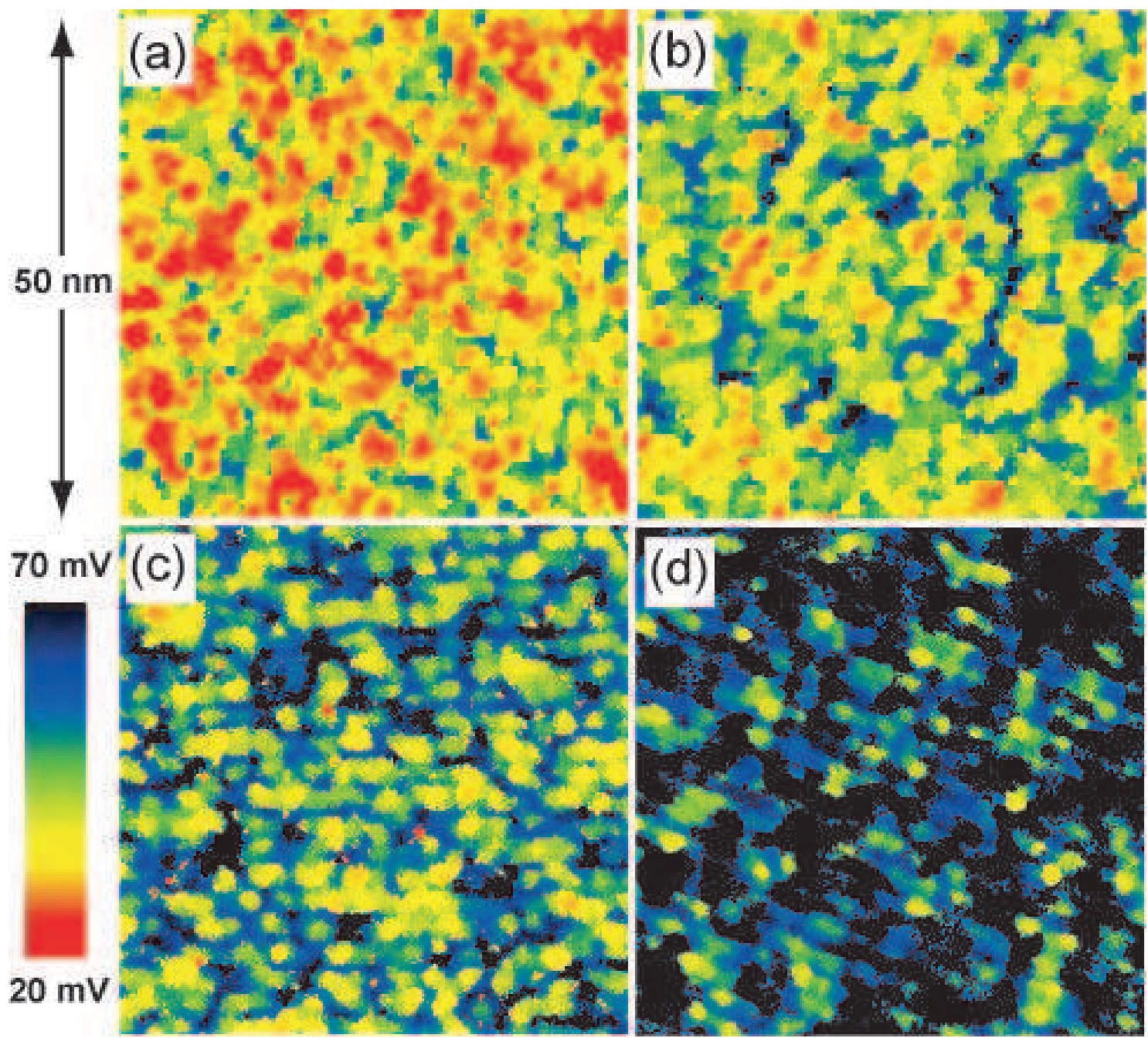
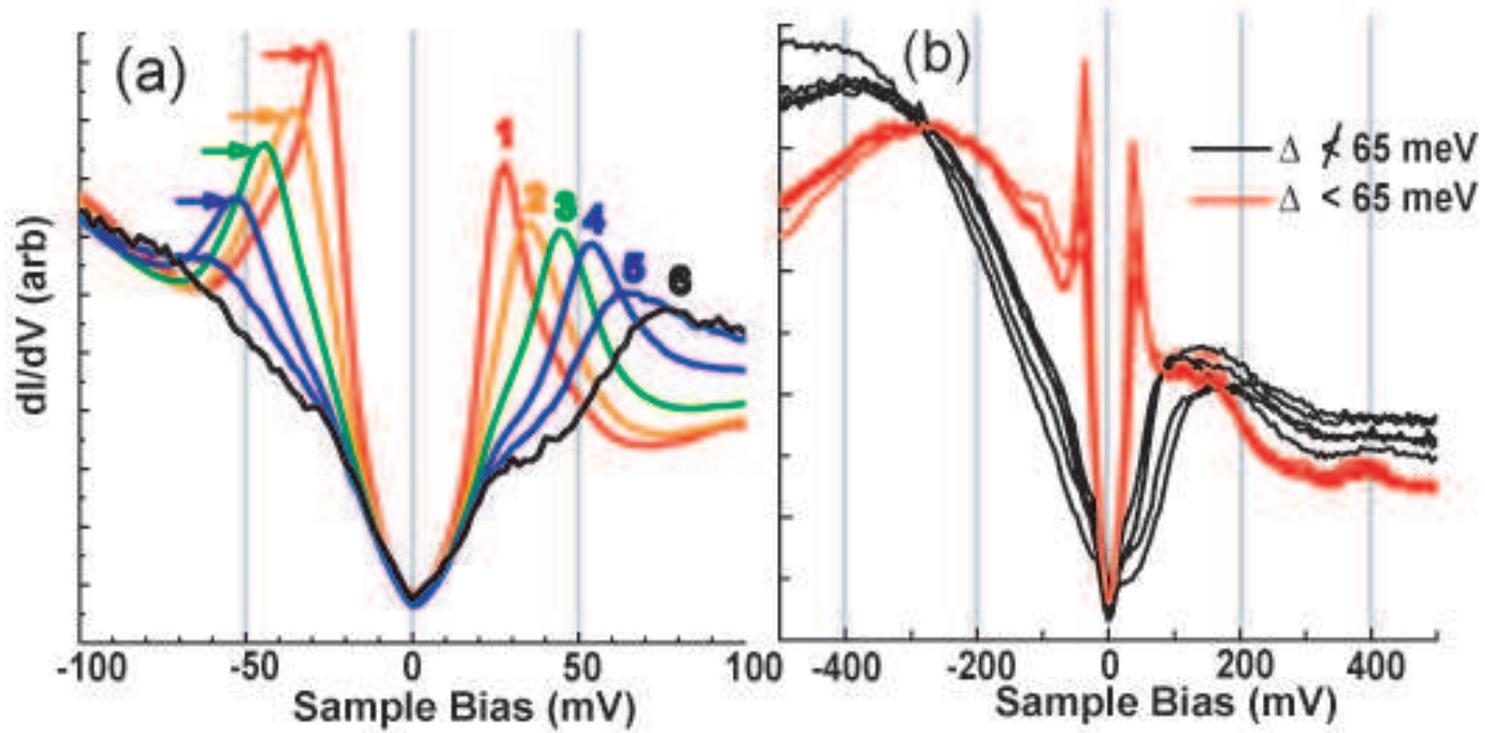
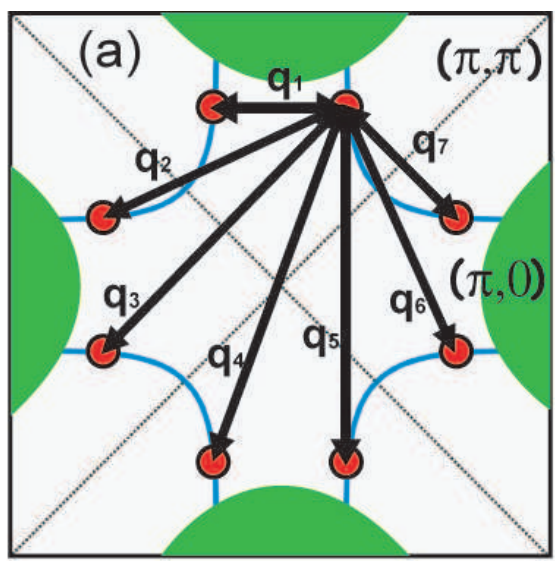
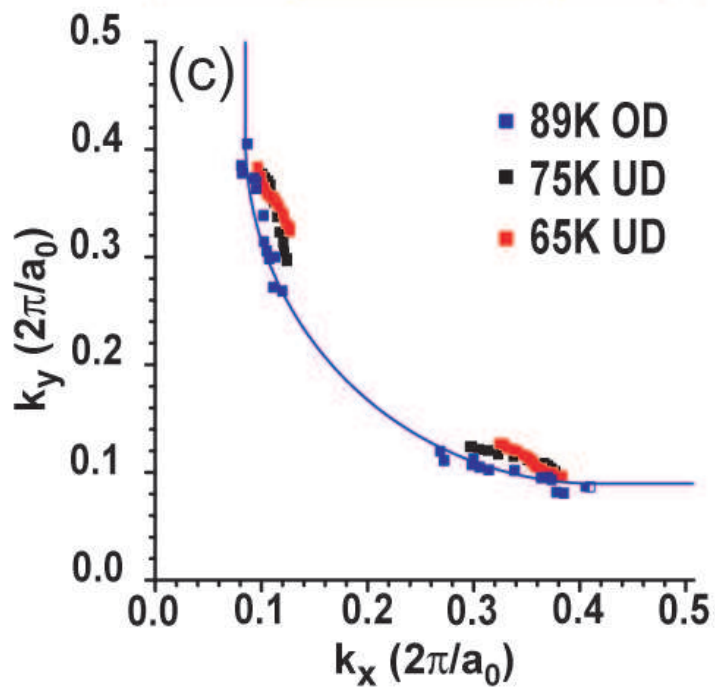
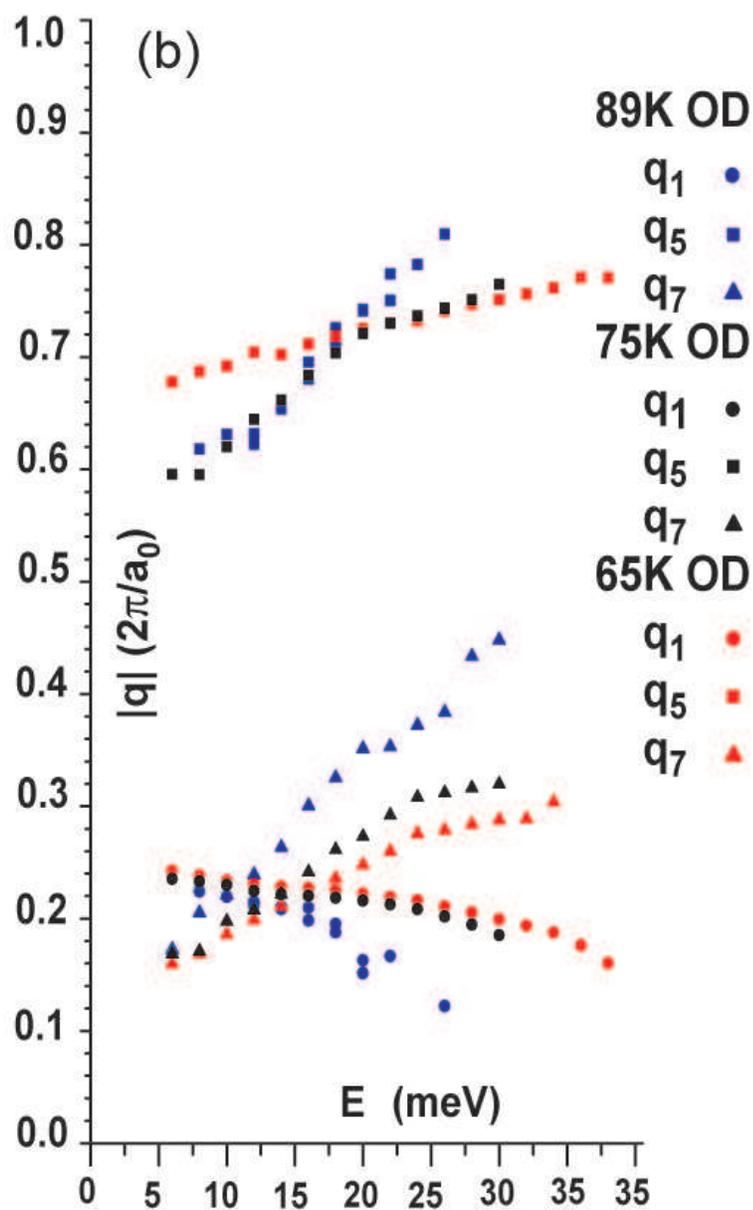
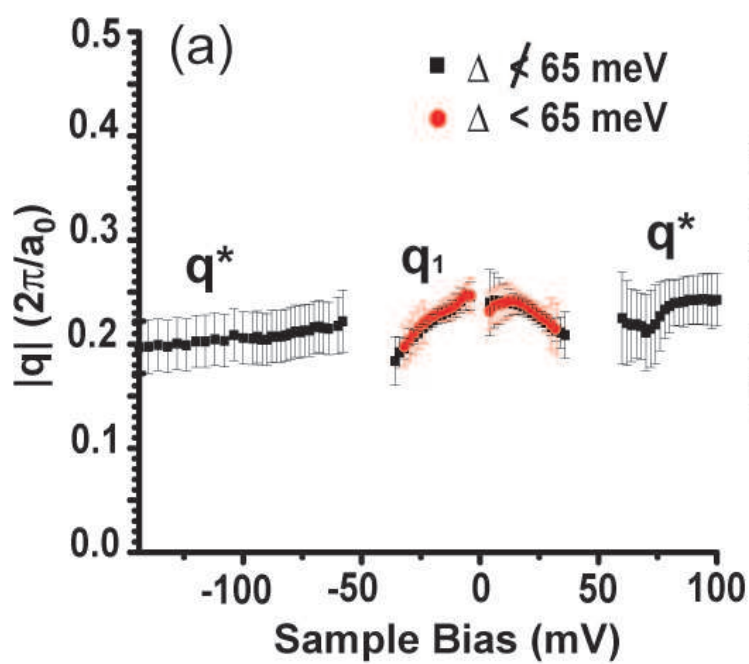
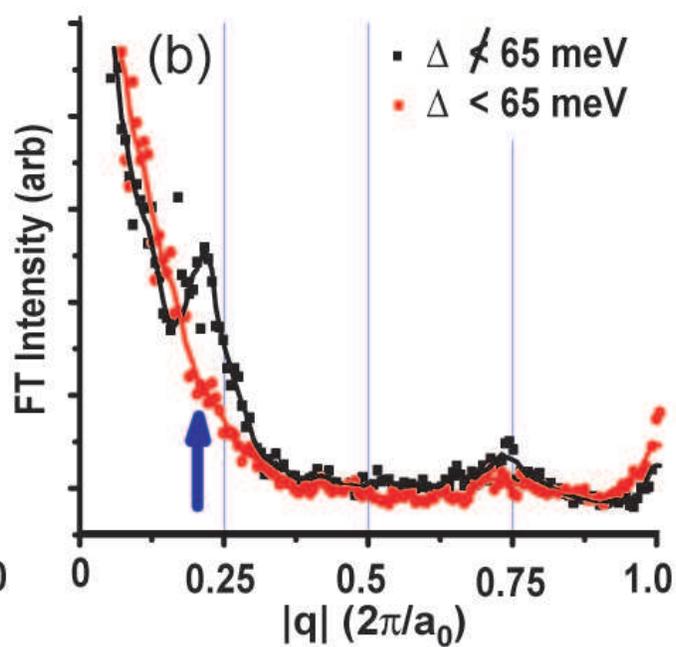
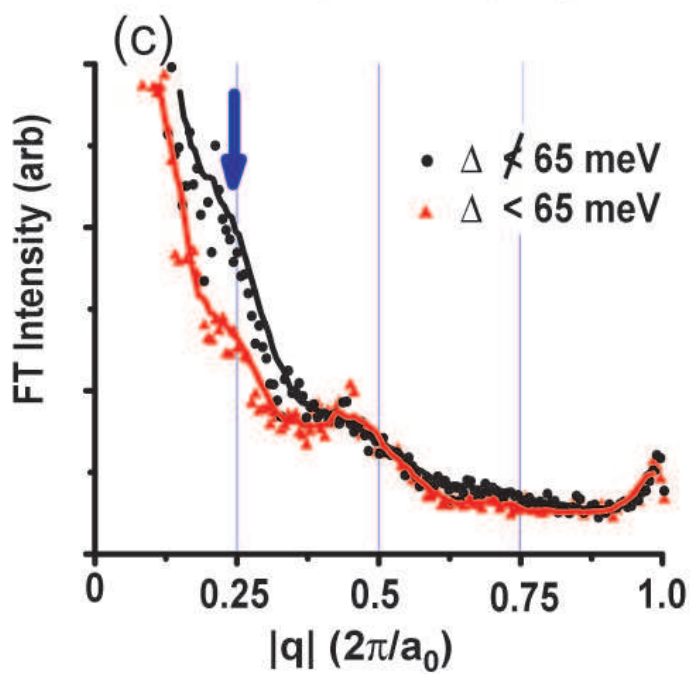
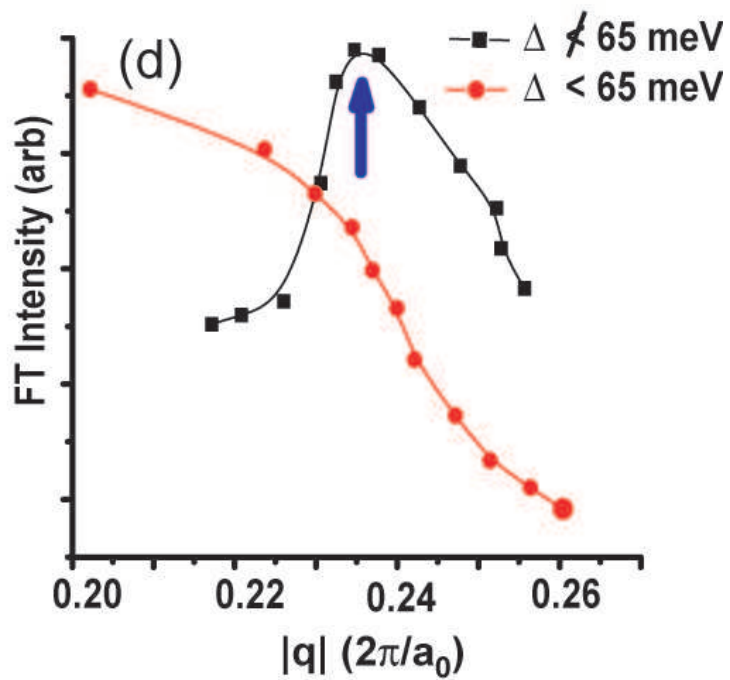
\end{document}